\documentclass[preprintnumbers,amsmath,amssymb]{revtex4}

\usepackage{color}
\usepackage{graphicx}
\usepackage{dcolumn}
\usepackage{bm}

\begin{document}


\title{Maximal attainable boost and energy of elementary  particles as a manifestation of the limit of localizability of elementary quantum systems}
\author{George Japaridze}
\affiliation{Department of Physics, Clark Atlanta University, Atlanta, Georgia, 30314, USA }


\date{\today}

\begin{abstract}
I discuss an upper bound on the boost and the energy of elementary particles. The limit is derived utilizing the core  principle of relativistic quantum mechanics stating that there is a lower bound for localization of an elementary quantum system and the assumption that when the localization scale reaches the Planck length, elementary particles are removed from the $S$-matrix observables.
The limits for the boost and energy, $M_{\mathrm{Planck}}/m$ and $M_{\mathrm{Planck}}c^{2}\approx\,8.6\cdot 10^{27}\,\mathrm{eV}$, are defined in terms of fundamental constants  and the mass of elementary particle and does not involve any dynamic scale. These bounds imply that the cosmic ray flux of any flavor may stretch up to energies of order $10^{18}$ GeV and will cut off around this value.
\end{abstract}

\maketitle

\section{Introduction}
This letter presents the scenario establishing the ultimate upper bound on the Lorentz boost and energy of elementary particles with a non-zero mass. The cosmic ray detectors measure higher and higher energies; recently the IceCube collaboration released data containing extra terrestrial neutrino events with $E_{\nu}\sim$ PeV ($10^{15}$ eV) \cite{icecube}. The region of energies probed is steadily increasing: cosmic ray experiments as HiRes, Telescope Area,  Auger report events with unprecedented energies of up to $10^{20}$ eV \cite{CR}. 

It is natural to ask whether there is no upper bound on Lorentz boosts/energies as it follows from the classical relativity. Accounting for the quantum feature of elementary particles and gravity may result in halting the increase of boost and energy of elementary particles. Though no deviation from Lorentz invariance is observed, which implies that space-time symmetry is described by a non compact group and that boosts and energies can acquire arbitrarily large values, there have been numerous attempts to modify the untamed growth of the boost and energy, see e.g. \cite{mat}, \cite{alan}.  Present note addresses this question.

The suggested mechanism for the maximal attainable boost and the energy of elementary particles is based on the fundamental concept of the limit of localizability of elementary quantum system in relativistic quantum mechanics \cite{nw}. This limit is combined with the conjecture that when the localization scale of the elementary particle reaches the critical value, defined by the maximum between the Planck length and Schwarzschild radius, it can not be observed as a quantum of an asymptotic {\it in, out} filed and thus elementary particle is not $S$-matrix observable any more.  For brevity, let us call this assumption the quantum hoop conjecture, since it  can be viewed as a quantum counterpart to the well-known hoop conjecture, which suggests that when the system is localized inside the volume of size of classical gravitational length, Schwarzschild radius, system undergoes gravitational collapse \cite{hoop}. Combining the quantum hoop conjecture with the limit of localizability of the elementary particle leads to the upper bound on Lorentz boost and energy of elementary particles.  Assuming that the mass of the elementary quantum system $m$ is less than Planck mass $M_{\mathrm{P}}$, this model predicts value of maximal attainable boost  $\Gamma^{\mathrm{max}}=M_{\mathrm{P}}/m$ and of maximal attainable energy $E^{\mathrm{max}}=M_{\mathrm{P}}c^{2}$.  When the boost reaches $\Gamma^{\mathrm{max}}$,  the contracted localization scale reaches the critical value and elementary particle is not  the $S$-matrix observable any more. The limits on boost and energy depend only on the fundamental constants and the mass and therefore can be considered as the ultimate bounds for the boost and energy of an elementary particle.

\section{Maximal attainable boost and maximal attainable energy}
To begin with, let us recall fundamental units. In this letter Planck mass $M_{\mathrm{P}}$ is defined as the value of a mass parameter at which the Compton wave length of a system with the mass $m$, $\lambda_{q}(m)=\hbar/mc$, is equal to the Schwarzschild radius, $\lambda_{\mathrm{gr}}(m)=2Gm/c^{2}$ (throughout a four dimensional space-time, no extra dimensions, is considered)
\begin{equation}
\label{Plank}
\lambda_{q}(M_{\mathrm{P}})\equiv{\hbar\over M_{\mathrm{P}}c}=\lambda_{\mathrm{gr}}(M_{\mathrm{P}})\equiv{2GM_{\mathrm{P}}\over c^{2}},
\end{equation}
where $\hbar$ is the Planck's constant, $G$ is the gravitational constant and $c$ is the speed of light. This results in 
\begin{equation}
\label{m1}
M_{{\mathrm{P}}}=\sqrt{{\hbar c\over 2 G}}\,\simeq\,1.5\cdot 10^{-8}\mathrm{kg}\simeq8.6\,\cdot\,10^{27}\,\mathrm{eV}/c^{2}.
\end{equation}
Planck length $\lambda_{{\mathrm{P}}}$ is defined as a Compton wave length of system with the Planck  mass, evidently coinciding with the Schwarzschild radius of a system with the Planck mass. From Eq. (\ref{Plank})  it follows
\begin{equation}
\label{plength}
\lambda_{{\mathrm{P}}}\,\equiv {\hbar\over M_{\mathrm{P}}c}\,=\,{2GM_{\mathrm{P}}\over c^{2}}\,=\,\sqrt{{2\,G\,\hbar\over c^{3}}}\simeq \,2.3\cdot 10^{-35}\,\mathrm{m}.
\end{equation}
These values differ by a factor of $\sqrt{2}$ from the ones established on a purely dimensional basis. Note the relation for a mass-independent Planck length made up from the mass-dependent $\lambda_{q}(m)$ and $\lambda_{\mathrm{gr}}(m)$:
\begin{equation}
\label{lll}
\lambda_{q}(m)\,\lambda_{\mathrm{gr}}(m)\,=\,\lambda^{2}_{{\mathrm{P}}}.
\end{equation}

We say that system is localized inside volume with a linear size $L_{0}$ when the probability of finding the system inside the ball of linear size $L_{0}$ is $1$. Observation is described by exchanging quanta of {\it in, out} fields between the localized system and the outside observer, separated from the system by distance much greater than $L_{0}$.

In classical physics the feature of rearranging the physical degrees of freedom when the localization scale reaches its critical value is  formulated by the hoop conjecture, which states that a  black hole forms whenever the amount of energy $mc^{2}$ is compacted inside a region that in no direction extends outside a circle of circumference (roughly) equal to $2\pi\lambda_{gr}(m)$  \cite{hoop}. In other words, no signal from the ball radius $a$ can reach external observer when $a<\lambda_{gr}(m)$. Originally, hoop conjecture was put forward for astrophysical bodies, macroscopic objects which can be reasonably described by classical theory of gravity \cite{hoop}. 

To establish the maximal attainable boost we introduce and utilize a quantum counterpart to the hoop conjecture, which we call throughout the quantum hoop conjecture. Quantum hoop conjecture states that whenever the localization scale of elementary quantum system approaches the Planck length $\lambda_{{\mathrm{P}}}$, the system is not observed as elementary particle, as the quantum  of  asymptotic {\it in, out} fields.
The idea that below the Planck length the notion of space-time and length  ceases to exists is not new, see \cite{mead}-\cite{sabine}; we just express it in terms of elementary particles, quanta of {\it in, out} fields.

In the solution of the Heisenberg equations of motion of quantum field theory  \cite{qft}
\begin{equation}
\label{HHH}
\langle \alpha|\hat{\Psi}|\beta \rangle = \langle \alpha|\sqrt{Z}\hat{\Psi}_{in,\,out}|\beta \rangle+\langle\alpha|{\hat R}|\beta \rangle,
\end{equation}
the term $\langle \alpha|\sqrt{Z}\hat{\Psi}_{in,\,out}|\beta \rangle$ describes an incoming/outgoing free particle, the quantum of asymptotic field and $\hat{R}$ stands for the rest of the solution, which in case of  a classical source $j(x)$ is $\hat{R}(x)=\int\,d^{4}y\,j(y)\triangle_{ret,\,adv}(x-y)$. In terms of quantum field theory, the quantum hoop conjecture is  translated into the statement that when the localization scale reaches the Planck length, the matrix element $\langle \alpha| \sqrt{Z}\hat{\Psi}_{in,\,out}|\beta\rangle$ vanishes,
\begin{equation}
\label{LLLKK}
\lim_{\Gamma\to\Gamma^{\mathrm{max}}}\,\langle \alpha(\Gamma)| \sqrt{Z}\hat{\Psi}_{in,\,out}|\beta(\Gamma)\rangle\,=\,0.
\end{equation}
Note that the flux of the incoming/outgoing particles and consequently, the $S$-matrix elements are defined by $\hat{\Psi}_{in,\,out}$; if matrix element of $\hat{\Psi}_{in,\,out}$ vanishes, corresponding $S$-matrix element vanishes as well \cite{qft}. According to quantum hoop conjecture 
the $S$-matrix element, corresponding to scattering of particle with the boost exceeding the maximal attainable boost, $\Gamma(m)\,>\,\Gamma^{\mathrm{max}}(m)$, should vanish.   The simplest way to realize this is to 
modify the expression for the operator of {\it in} field:
\begin{equation}
\label{psi1}
\hat{\Psi}_{in}\,=\,\sum_{{\bf k}}\,{a_{in}({\bf k})\,e^{-ikx}+a^{\dagger}_{in}({\bf k})\,e^{ikx}\over 2k_{0}}\,\to\,\sum_{{\bf k}}\,{\widetilde{a}_{in}({\bf k})\,e^{-ikx}+\widetilde{a}^{\dagger}_{in}({\bf k})\,e^{ikx}\over 2k_{0}}, 
\end{equation}
where $k_{0}=\sqrt{{\bf k}^{2}+m^{2}}$, $\widetilde{a}_{in}({\bf k})\equiv a_{in}({\bf k})\,\Theta(E^{\mathrm{max}}\,-\,k_{0})$,  $a_{in}({\bf k}),\,a^{\dagger}_{in}({\bf p})$ are the annihilation and creation operators of elementary particle (quantum of {\it in} field) satisfying the usual commutator relation $[a_{in}({\bf k}),\,a^{\dagger}_{in}({\bf p})]=\delta({\bf k}-{\bf p})$, $\Theta(E^{\mathrm{max}}\,-\,k_{0})$ is the Heaviside step function and $E^{\mathrm{max}}$ is the energy of particle mass $m$ boosted with $\Gamma^{\mathrm{max}}(m)$. 
From (\ref{psi1}) it follows that the matrix element of asymptotic field $\hat{\Psi}_{in}(x)$ between the vacuum and the one particle state $|k\rangle=a^{\dagger}_{in}(k)|0\rangle$ is
\begin{equation}
\label{inset3}
\langle 0|\hat{\Psi}_{in}(x)|k\rangle\,\sim\,\sqrt{Z}\,e^{ikx}\,\Theta(E^{\mathrm{max}}\,-\,k_{0}),
\end{equation}
which vanishes when $k_{0}\,\geq \,E^{\mathrm{max}}$. As far as $k_{0}\leq E^{\mathrm{max}}$, i.e. when $\Gamma\leq \Gamma^{\mathrm{max}}(m)$, the vacuum-one particle matrix element of an {\it in, out} field exists, i.e.  a particle can be observed as a quantum of $\hat{\Psi}_{in,\,out}$ and the flux and the $S$-matrix exist.  Phenomenological condition (\ref{psi1}) has to be derived from future theory of quantum gravity at a regime when an elementary particle having energy of the order of maximal attainable energy $E^{\mathrm{max}}$ in some reference frame would presumably interact with the degrees of freedom of the unknown underlying theory of quantum gravity. 

\noindent Eqs. (\ref{psi1}) and (\ref{inset3})  are written in a preferred reference frame.  As any other scenario speaking of the maximum attainable boost and energy and thus postulating the existence of preferred reference frame, the present model also requires the existence of the reference frame to which the maximum boost is compared. In this work, motivated by the cosmological considerations, for such a reference frame the cosmic rest frame is chosen. This is the reference frame where the cosmic microwave background (CMB) is at rest, and its temperature is homogeneous and is 2.73$K$. 
The Earth rest reference frame moves relative to CMB with peculiar velocity $\sim 370$ km/s $\approx 0.0012\,c$, as is follows from CMB dipole anisotropy measurements \cite{earth}. Because of the low value of  $\Gamma_{\mathrm {Earth-CMB}}\sim 1$, with a good approximation the Earth rest reference frame can be identified with the preferred reference frame, the one where as $\Gamma\,\rightarrow\,\Gamma^{\mathrm{max}}$, no elementary particle can be observed.

Maximal attainable boost for the elementary particle with mass $m$, $\Gamma^{{\mathrm max}}(m)$, is derived from the requirement that the Lorentz-contracted localization scale is still larger than either Schwarzschild radius or Planck length and therefore, elementary particle is still observable as a quantum of {\it in, out} fields.  From the classical special relativity it follows that when boosted with $\Gamma$, the localization scale is spatially contracted and becomes $L=L_{0}/\Gamma$. This relation holds in a relativistic quantum mechanics as well - position operator in relativistic quantum mechanics acquires overall factor $1/\Gamma$ when the reference frame is boosted with $\Gamma$ \cite{nw1}. From the requirement that the Lorentz-contracted  size of the system is still larger than the threshold value $L_{\mathrm{thr}}$, given by either gravitational radius or Planck scale
\begin{equation}
\label{boost0}
L={L_{0}\over \Gamma}\,\geq\,L_{\mathrm{critical}}\equiv \mathrm{max}(\lambda_{\mathrm{P}},\,\lambda_{\mathrm{gr}}(m)),
\end{equation}
it follows that
\begin{equation}
\label{boost1}
\Gamma\,\leq\, \Gamma^{\mathrm{max}}(m)={L_{0}\over L_{\mathrm{critical}}}\,=\,{L_{0}\over \lambda_{\mathrm{q}}(m)}\,{\lambda_{\mathrm{q}}(m)\over L_{\mathrm{critical}}}.
\end{equation}
We utilize the well-known fact that in the framework of relativistic quantum mechanics, a particle at rest can not be localized with accuracy better  than its Compton wave length $\lambda_{\mathrm{q}}(m)$, i.e. min$(L_{0})=\lambda_{\mathrm{q}}(m)$   \cite{nw}, \cite{qft}.  
Then the inequality (\ref{boost1}) turns into
\begin{equation}
\label{boost2}
\Gamma\,\leq\, \Gamma^{\mathrm{max}}(m)\,=\,{\lambda_{\mathrm{q}}(m)\over L_{\mathrm{thr}}}.
\end{equation}
We assume that the mass of the elementary quantum system is bounded from above,  $m\leq M_{\mathrm{P}}$. This constraint is realized as an inequality ordering spatial scales as follows
\begin{equation}
\label{scales}
\lambda_{\mathrm{gr}}(m)\leq \lambda_{\mathrm{P}}\leq \lambda_{\mathrm{q}}(m),
\end{equation}
and consequently, $L_{\mathrm{critical}}=\mathrm{max}(\lambda_{\mathrm{P}},\,\lambda_{\mathrm{gr}}(m))=\lambda_{\mathrm{P}}$. Therefore, the maximal attainable boost for a particle with mass $m\leq M_{\mathrm{P}}$ is
\begin{equation}
\label{boost6}
\Gamma^{\mathrm{max}}(m)={\lambda_{\mathrm{q}}(m)\over \lambda_{\mathrm{P}}}={\lambda_{\mathrm{P}}\over \lambda_{\mathrm{gr}}(m)}={M_{\mathrm{P}}\over m}.
\end{equation}

The hoop conjecture which may serve as a physical insight for the rearrangement of the space of physical degrees of freedom, appeals to the Schwarzschild radius $\lambda_{\mathrm{gr}}$, as the minimal localization length. Connection between Planck length and gravity effects was established long ago, and it states that in presence of gravity it is impossible to measure the position of a particle with error less than $\lambda_{\mathrm{P}}$  \cite{mead}, \cite{garay}, \cite{carlo}. We have assumed that similar to the hoop conjecture, suggesting that object collapses into black hole  when localized in area with size less than classical gravitational radius, $2Gm/c^{2}$, elementary particle is removed from $S$-matrix observables when localized in volume with size less than $\lambda_{\mathrm{P}}$, the latter determined by both $G$ and $\hbar$.  When the mass of elementary particle approaches $M_{\mathrm{P}}$, as seen from Eq. (\ref{lll}), the Planck length and the Schwarzschild radius coincide
\begin{equation}
\label{joe}
\lambda_{\mathrm{P}}\,=\,\lambda_{\mathrm{gr}}(M_{\mathrm{P}})
\end{equation}
i.e. the quantum and the classical hoop conjectures merge into the same supposition.

According to Eq. (\ref{boost6}), the value of maximal attainable boost varies from particle to particle, e.g. $\Gamma^{\mathrm{max}}(\mathrm{proton})=M_{\mathrm{P}}/m_{\mathrm{proton}}\approx 9.2\cdot 10^{14}$, $\Gamma^{\mathrm{max}}(\mathrm{electron})\approx 1.7\cdot 10^{22}$. Not so for the maximal attainable energy: $E^{\mathrm{max}}$ is the same for all particles and is given by the Planck energy $E_{\mathrm{P}}\equiv M_{\mathrm{P}}c^{2}$:
\begin{equation}
\label{maxe}
E^{\mathrm{max}}=\Gamma^{\mathrm{max}}(m)\,mc^{2}\,=\,M_{\mathrm{P}}c^{2}\approx\,8.6\cdot\,10^{27}\,\mathrm{eV}
\end{equation}
It is important to note that the above results are obtained and are valid for the systems with $m\leq M_{\mathrm{P}}$ which we consider as elementary particles, i.e. quanta of {\it in, out} fields.

The disappearance of {\it in, out} fields from the solution of Heisenberg equations of motion, in other words, removing elementary particles from the $S$-matrix observables when $\Gamma\,>\, \Gamma^{\mathrm{max}}$, seems to violate the $S$-matrix unitarity. Indeed, asymptotic completeness, according to which the {\it in} and {\it out} states span the same Hilbert space, which is also assumed to agree with the Hilbert space of interacting theory \cite{qft} 
\begin{equation}
\label{Hilb}
{\cal H}_{in}\,=\,{\cal H}_{out}\,={\cal H}_{\mathrm{interacting}}
\end{equation}
is not satisfied. 
Condition of  asymptotic completeness is not trivial already in the framework of standard quantum field theory: if particles can form bound states, the structure of space of states is modified, and the $S$-matrix unitarity is restored only after bound states are accounted for in the unitarity condition \cite{qft}. Accounting for gravity brings in another reasoning for the violation of the $S$-matrix unitarity. It has previously been observed that {\it in} and {\it out} states, which are related by unitary transformation, can not be be defined in the presence of an arbitrary metric \cite{parker}. 
Applying the quantum hoop conjecture to elementary particles drives this observation to the extreme, stating that as soon as the localization region becomes smaller than $\lambda_{\mathrm{P}}$, the space of physical degrees of freedom is rearranged and elementary particles, quanta of {\it in, out} fields are not observables any more. In this case, the unitarity condition has to be formulated not in terms of elementary particles, but in terms of new physical degrees of freedom of quantum gravity, task which is beyond the scope of this letter. 

As for the predictions of a suggested scenario, the only clear one is the cut-off of beam of particles/flux of cosmic rays at limiting value $E_{\mathrm{P}}\sim 10^{18}$ GeV, energy, which is not accessible by  modern accelerators and the cosmic rays observatories. Up to  $E_{\mathrm{P}}$, i.e. up to $\Gamma^{\mathrm{max}}$ when the localization region is still larger than the Planck length, physical degrees of freedom, observables, are elementary particles.  When $E\geq E_{\mathrm{P}}$, elementary particles are removed from observables (in full analogy of collapsing system into a black hole according to a hoop conjecture); thus cosmic ray flux should vanish when $E\rightarrow 10^{18}$ GeV. As mentioned  above, maximum attainable boost varies from particle to particle and is $M_{\mathrm{P}}/m$, but maximum attainable energy for any type of elementary particle is the same $E_{\mathrm{P}}$. These bounds are ``kinematical'' in a sense that no dynamic scale related with any particular interaction is involved in establishing the bound. Of course, some concrete conditions may alter the maximal observed    energy. e.g the well-known GZK limit on the energy of cosmic rays from distant sources, caused by the existence of the omnipresent target - cosmic microwave background \cite{gzk}.  However the presented bounds are ultimate, derived from the quantum hoop conjecture and the basic principles of relativistic theory of quantum systems; in other words statement is that independently of dynamics the kinematic parameters describing elementary particles can not exceed these bounds. 
 
This is in contrast with the scenario for the  boost and energy cut-off which was recently put forward \cite{learned}. In \cite{learned} it is suggested that the maximum attainable boost and maximum attainable energy for the neutrino are
\begin{equation}
\label{learn}
\Gamma^{\mathrm{max}}_{\nu}={M_{\mathrm{P}}\over M_{\mathrm{weak}}};\quad E^{\mathrm{max}}_{\nu}=m_{\nu}{M_{\mathrm{P}}\over M_{\mathrm{weak}}}
\end{equation}
where $M_{\mathrm{P}}$ is a Planck scale and $M_{\mathrm{weak}}$ is a scale of weak interactions ($\sim 100$ GeV). The main point of work \cite{learned} is that the upper bound on energy is defined by  weak scale; it follows from (\ref{learn})  that neutrino spectrum cuts off at energies $\sim$ few PeV. Our prediction is  $\Gamma^{\mathrm{max}}_{\nu}=M_{\mathrm{P}}/m_{\nu}$, i.e. much higher than one from Eq. (\ref{learn}).  Regard neutrino energies, as an example, we quote the estimate from analysis of energetics of gamma-ray bursts -  maximum neutrino energies may reach $10^{16}-5\cdot 10^{19}$ eV \cite{grb}. This value does not contradict  Eq. (\ref{maxe}) - $E^{\mathrm{max}}_{\nu}= 8.6\cdot 10^{27}$ eV, and exceeds the upper bound of a few PeV on neutrino energies suggested in \cite{learned}. 

\section{Discussion}
We have combined: the lower limit of localizability of an elementary quantum system, Lorentz contraction, and the quantum hoop conjecture to derive the upper bound on Lorentz boost and energy for massive particles.  When the upper bound is reached, elementary particles - the quanta of asymptotic {\it in, out} fields disappear from a spectrum of $S$-matrix observables, presumably replaced by the local physical degrees of freedom of quantum gravity. In derivation of this upper bound, we used the property of Lorentz contraction, i.e. validity of the theory of relativity applied to elementary particles up to $\Gamma^{\mathrm{max}}$ is assumed.  
Though the limiting values of boost and energy are Lorentz invariant, assigning the physical meaning to  $\Gamma^{\mathrm{max}}$ and $E^{\mathrm{max}}$ implies the existence of a preferred reference frame. In this work it is postulated that the preferred reference frame is the CMB rest reference frame. Since the Earth rest reference frame almost coincides with the CMB rest frame, we predict that in the Earth rest reference frame  the cosmic ray spectrum continues all the way till the Planck energy $\sim 10^{18}$ GeV, where the cut-off of the flux occurs.

Lastly, let us note that as it is assumed that the boost is bounded from above, i.e. for a massive particle the (classical) limit $v=c$ can not be reached, the classical hoop conjecture remains intact. This is because the metric remains of that of boosted Schwarzschild space-time and can not be approximated by a plane impulsive gravitational wave as it happens for states moving with $c$ \cite{light}.

I thank V.A. Petrov for fruitful discussions and pointing out to me works regarding boosted gravitational fields of massless particles \cite{light}.




\begin{thebibliography}{100}
\bibitem{icecube}
M. G. Aartsen et al. [IceCube Collaboration],  {\it Phys. Rev. Lett.}, {\bf 113}, 101101, (2014);  arXiv:1405.5303. 
\bibitem{CR}
P. Blasi, Plenary talk at at the 33rd International Cosmic Ray Conference, 2013, Rio de Janeiro, Brazil; 	arXiv:1312.1590 [astro-ph.HE].
\bibitem{mat}
D. Mattingly, 
{\it Living Rev. Rel.}, {\bf 8},  5, (2005); arXiv:gr-qc/0502097; 
S. Liberati, {\it Class. Quantum Grav.}, {\bf 30}, 133001, (2013).
\bibitem{alan}
A. Kostelecky and N. Russell, {\it Rev. Mod. Phys.}, {\bf 83:11}, 2011.
\bibitem{nw}
T.D. Newton and E.P. Wigner, {\it Rev. Mod. Phys.}, {\bf 21}, 400, (1949).
\bibitem{hoop} 
K.S. Thorne, 
in {\it J.R. Klauder, Magic Without Magic}, Freeman, S. Francisco, 231, (1972).
\bibitem{mead}
C. A. Mead, {\it Phys. Rev.}, {\bf 135}, B849, (1964), {\it Phys. Rev.}; {\bf 143}, 990, (1966).
\bibitem{garay}
L. Garay, Int. J. Mod. Phys. {\bf A10}, 145, (1995).
\bibitem{carlo}
S. Doplicher, K. Fredenhagen and J.E. Roberst, {\it Commun. Math. Phys.}, {\bf 172}, 187, (1995); C. Rovelli and L. Smolin, Nucl. Phys. {\bf B442}, 593 (1995).
\bibitem{sabine}
S. Hossenfelder, {\it  Living Rev. Rel.}, in {\bf 16}, 2 (2013); arXiv:0806.0339v2 [gr-qc] .
\bibitem{qft}
S. Weinberg, {\it The Quantum Theory of Fields}, Campridge University Press, 2005.
\bibitem{earth}
C. L. Bennett et al., {\it  Astrophys. J. Supp.} {\bf 148}, 1 (2003).
\bibitem{nw1}
A.H. Monahan and M. McMillan, {\it Phys. Rev. A}, {\bf 56}, 2563, (1997).
\bibitem{parker}
L. Parker, {\it Phys. Rev.}, {\bf 183}, 1057 (1969); S. A. Fulling, L. Parker and B. L. Hu, {\it Phys. Rev.} {bf 10}, 3905 (1974).
\bibitem{gzk}
K. Greisen, {\it Phys. Rev. Lett}, {\bf 16}, 748 (1966); G. Zatsepin and V. Kuzmin, {\it JETP Lett.}, {\bf 4}, 78 (1966).
\bibitem{learned}
J.G. Learned and T.J. Weiler, 	arXiv:1407.0739 [astro-ph.HE].
\bibitem{grb}
P. Kumar and B. Zhang, {\it Phys. Rep.}, {\bf 561}, 1, (2015).
\bibitem{light}
W.B. Bonnor, {\it Commun. math. Phys.}, {\bf 13}, 163 (1969); P.C. Aichelburg and R.U. Sexl, {\it Gen. Rel. Grav.}, {\bf 2}, 303 (1971).
\end{thebibliography}
\end{document}